\documentclass[aip,preprint]{revtex4-2}

\usepackage{times}
\usepackage{array}
\usepackage{longtable}
\usepackage{float}
\usepackage{amsmath}
\usepackage{bm}
\usepackage{graphicx}
\usepackage{cancel}
\usepackage{amssymb}
\usepackage{xcolor}
\usepackage{hyperref}
\hypersetup{bookmarks=true,colorlinks=true, linkcolor=blue,linktoc=all}
\usepackage[FIGTOPCAP]{subfigure}

\makeatletter
\usepackage{fancyhdr}
\usepackage{babel}
\makeatother

\begin{document}

\title{A Physics-Informed Neural Network for Solving the Quasi-static Magnetohydrodynamic Equations}

\author{Jonathan S. Arnaud}
\email{jsarnaud@lanl.gov}
\affiliation{Nuclear Engineering Program, Department of Materials Science and Engineering, University of Florida, Gainesville, FL 32611, United States of America}
\affiliation{Theoretical Division, Los Alamos National Laboratory, Los Alamos, NM 87545, United States of America}

\author{Christopher J. McDevitt}
\affiliation{Nuclear Engineering Program, Department of Materials Science and Engineering, University of Florida, Gainesville, FL 32611, United States of America}

\author{Golo Wimmer}
\affiliation{Theoretical Division, Los Alamos National Laboratory, Los Alamos, NM 87545, United States of America}

\author{Xian-Zhu Tang}
\affiliation{Theoretical Division, Los Alamos National Laboratory, Los Alamos, NM 87545, United States of America}
\date{\today}

\begin{abstract}
A physics-informed neural network (PINN) is developed, for the first time, to learn the time-dependent quasi-static magnetohydrodynamic (MHD) equations in axisymmetric tokamak geometry, without any experimental or synthetic data. The initial study considered an ITER-like tokamak and found that a PINN, after careful treatment, was capable of learning the solution to the MHD system and predict a vertically displacing plasma, where general agreement with ground truth simulation was observed. The proof-of-principle demonstration highlights the potential of physics-constrained deep learning to learn complex plasma behavior.
\end{abstract}
\maketitle

In magnetic confinement devices known as tokamaks, magnetohydrodynamics (MHD) provides an efficient means of treating the motion of the bulk plasma during a tokamak disruption. For an elongated plasma, the plasma will often become vertically unstable, resulting in a vertical displacement event (VDE) that can result in unwanted damage on device components. Due to planned next-generation devices anticipating potentially substantial damage from unmitigated disruptions~\cite{bandyopadhyay2025mhd}, in addition to operating in regimes inaccessible by present day experiments, an efficient and accurate simulation framework is required to design operating scenarios that prevent or mitigate the unwanted damage from tokamak disruptions. This has motivated the use of machine learning to accelerate scientific computing, due to the rapid inference time (typically microseconds to milliseconds) and hardware acceleration from graphical processing units (GPUs).

Physics-informed machine learning~\cite{karniadakis2021physics} methods in particular, such as physics-informed neural networks~\cite{raissi2019physics} directly embed partial differential equations (PDEs) into the learning algorithm to enrich incomplete or sparse data sets, and has recently been applied to equilibrium reconstruction in tokamaks~\cite{rossi2023potential} and stellarators~\cite{thun2026improving}. An attractive feature of PINNs is that data can be omitted in the learning process, which allows the PINN to directly learn the solution to a PDE~\cite{dissanayake1994neural, van1995neural, lagaris1998artificial}. Indeed, one
of the earliest works on using PINNs to learn a PDE solution was by van Milligen and colleagues in 1995~\cite{van1995neural} and 1997~\cite{van1997solving} for tokamak and stellarator equilibriums, respectively. Another attractive feature of PINNs is the seamless ability to learn parametric PDE solutions and serve as rapid high-fidelity surrogate models (See Refs.~\cite{luo2025parametric,jang2024grad} for examples). An open question thus becomes apparent, being if PINNs can be used to learn the MHD solution for describing the plasma behavior during a VDE across a broad range of scenarios, which would allow for rapid iteration in designing an operating scenario that mitigates or prevents unwanted damage.
\begin{figure}
\centering
\includegraphics[height=0.5\textwidth]{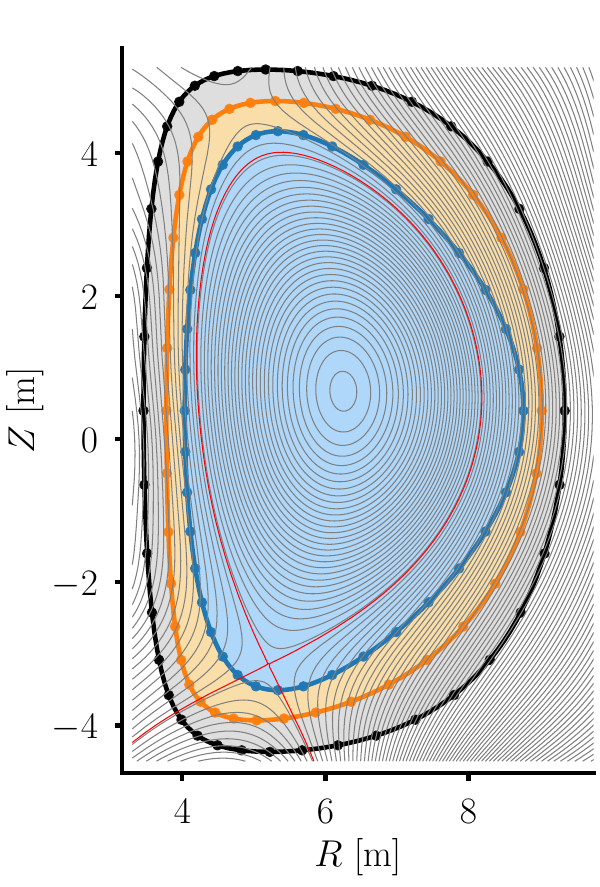}
\caption{ITER-like equilibrium, where the grey contours correspond to $\psi$, and the shaded regions correspond to the vacuum chamber (blue), blanket (orange), and vessel (black) regions.}
\label{fig:1}
\end{figure}

In this letter we investigate if a PINN can learn the solution to a set of time dependent MHD equations describing a VDE without any synthetic or experimental data, which to the best of our knowledge, is the first time this is done. As a first step, we will determine if a PINN can learn the MHD solution for a single set of plasma parameters. Beginning with the MHD formulation, we consider the scenario where a disrupted plasma has lost the majority of its thermal energy and becomes vertically unstable, leading to motion towards the walls of the device. For this slow resistive evolution of the plasma a suitable description is quasi-static resistive MHD, which assumes the limit of the plasma being ``force-free'', due to the low temperatures and thermal pressures typically found in tokamak disruptions, and can be written as~\cite{kiramov2018force}
\begin{equation}
\frac{\partial\mathbf{B}}{\partial t} = \nabla\times (\mathbf{u}\times\mathbf{B}) - \nabla\times(\eta\mathbf{j}),
\label{eq:1}
\end{equation}
\begin{equation}
0 = \mathbf{j}\times\mathbf{B} + \frac{1}{Re}\nabla^2\mathbf{u},
\label{eq:2}
\end{equation}
where $\mathbf{B}$ is the magnetic field, the plasma flow velocity is $\mathbf{u}$, the resistivity is $\eta$, the current density is $\mathbf{j}$, time is $t$, and $Re$ is the Reynolds number. We note that the toroidal component of $\mathbf{u}$ will be omitted, due to the dominant form of transport arising from the poloidal flow. We will consider a cylindrical coordinate system $(R,\varphi,Z)$ that assumes symmetry in the toroidal angle $\varphi$. The resulting 2D-MHD system can be straightforwardly solved by prescribing an initial condition for $\mathbf{B}$ and the appropriate boundary conditions. The initial condition is acquired by solving the free-boundary Grad-Shafranov problem, which is done through the use of the Free-boundary Grad-Shafranov Newton-Krylov Evolve (FreeGSNKE) code~\cite{amorisco2024freegsnke} for an ITER-like configuration that contains 15 MA of plasma current and an on-axis magnetic field of 5 T. Given that the equilibrium obtained by FreeGSNKE will not perfectly satisfy force-balance, a rapid flow will develop to drive the system to be force-free. To avoid having the PINN learn this artificially rapid flow, we will allow the system to relax until force-balance is restored, and pass the corresponding equilibrium as the initial condition to the PINN. For the boundary conditions, we consider an ITER-like geometry consisting of three regions, being the vacuum chamber, the blanket region, and the vessel structure, where the boundaries can be conveniently expressed by $R(\theta) = R_c + R_c\epsilon\cos(\theta + \alpha(\delta)\sin\theta), Z(\theta) = Z_c + R_c\epsilon\kappa\sin(\theta + \zeta\sin2\theta)$, and $\alpha(\delta) = \sin^{-1}(\delta)$ is taken to piecewise, such that $\delta = \delta_\mathrm{upper}$ for $0 \leq \theta \leq \pi$ and $\delta = \delta_\mathrm{lower}$ for $< \theta 2\pi$. The shape parameters consist of the elongation $\kappa$, inverse aspect ratio $\epsilon$, and triangularity $\delta$.  A schematic of the geometry and initial equilibrium is shown in Figure~\ref{fig:1}, where we have chosen $ R_{c} = 6.4, Z_{c} = 0.4, \epsilon_\mathrm{vacuum} = 0.37, \epsilon_\mathrm{blanket} = 0.41, \epsilon_\mathrm{vessel} = 0.46, \kappa_\mathrm{vacuum} = \kappa_{blanket} = 1.65, \kappa_{vessel} = 1.62, \delta_\mathrm{upper,vacuum} = 0.45, \delta_\mathrm{upper,blanket} = \delta_\mathrm{upper,vessel} = 0.42, \zeta_\mathrm{vacuum} = -0.03, \zeta_\mathrm{blanket} = 0.15, \zeta_\mathrm{vessel} = 0.2$. With the prescribed geometry we can impose boundary conditions, which are that $\mathbf{u}$ is taken to vanish on the vacuum chamber boundary, and a perfectly conducting boundary at the edge of the vessel region $(\mathbf{E}\times\mathbf{n} = 0)$.

While a PINN could be deployed to learn the system given by Equations~\ref{eq:1} and \ref{eq:2}, additional constraints would need to be imposed, such as $\nabla\cdot\mathbf{B} = 0$, which would not be guaranteed. Instead, the magnetic vector potential $\mathbf{A} = \nabla\times\mathbf{B}$ will be utilized, which satisfies $\nabla\cdot\mathbf{B}=0$ by construction. Taking the axisymmetric magnetic field to be $\mathbf{B} = g\nabla\varphi +  \nabla\varphi\times\nabla\psi$, where $g = RB_\varphi$, the poloidal flux function $\psi = RA_\varphi$ is related to the poloidal flux by $\Psi = \int_0^{2\pi}RA_\varphi\mathrm{d}\varphi$, we can rewrite Equations~\ref{eq:1} and ~\ref{eq:2} as
\begin{equation}
\frac{\partial\psi}{\partial t} = \frac{R^2}{S}\left[\nabla\cdot\left(\frac{1}{R^2}\nabla\psi\right)\right] - \mathbf{u}\cdot\nabla\psi,
\label{eq:3}
\end{equation}
\begin{equation}
\frac{\partial g}{\partial t} = R^2\nabla\cdot\left[\frac{(\nabla g/S) - \mathbf{u}g}{R^2}\right],
\label{eq:4}
\end{equation}
\begin{equation}
0 = R^2\nabla\cdot\left(\frac{\nabla\psi}{R^2}\right)\frac{\partial\psi}{\partial R} + g\frac{\partial g}{\partial R} + \frac{1}{Re}(\nabla^2\mathbf{u})_R,
\label{eq:5}
\end{equation}
\begin{equation}
0 = R^2\nabla\cdot\left(\frac{\nabla\psi}{R^2}\right)\frac{\partial\psi}{\partial Z} + g\frac{\partial g}{\partial Z} + \frac{1}{Re}(\nabla^2\mathbf{u})_Z,
\label{eq:6}
\end{equation}
where $S$ is the Lundquist number, and we have performed the following normalizations: $S = S_0\tilde{S}, R = L_0\tilde{R}, Z = L_0\tilde{Z}, B = B_0\tilde{B}, t = S\tau_a\tilde{t}, \mathbf{u} = S\tau_a\mathbf{u}/L_0$, which yields $\psi = L^2_0B_0\tilde{\psi} \  \mathrm{ and } \  g = L_0B_0\tilde{g}$, where $\tau_a = L_0/v_a$ is the Alfv\'en speed timescale, we have chosen $L_0 = 2 \mathrm{m}$ and $B_0 = 5 \mathrm{T}$, and the reference Lundquist number will be taken to be the corresponding value in the vacuum chamber region. We choose the following Lundquist numbers: $S_\mathrm{vacuum} = 10^{5}, S_\mathrm{blanket} = 10^4, S_\mathrm{vessel} = 6 \times 10^5$, and $Re = 10$. Finally, the corresponding boundary conditions are $\partial\psi/\partial t = 0$ and $\nabla g \cdot \mathbf{n}=0$ at the edge of the vessel region. We note that while the discontinuities of $S$ in Equation~\ref{eq:4} will require conditions that enforce continuity on the interfaces, since a single PINN model will be used for the whole domain, these conditions will be automatically satisfied. Finally, Equations~\ref{eq:5} and~\ref{eq:6} are only valid inside the vacuum chamber where the plasma is present.   
\begin{figure}
\includegraphics[width=0.6\textwidth]{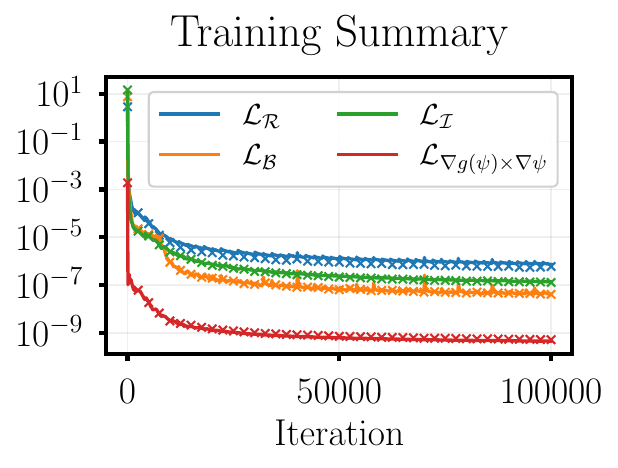}
\caption{Training (solid curves) and test (x markers) PINN loss.}
\label{fig:2}
\end{figure}

Deploying a PINN to learn a solution for complex systems typically requires special treatment, where the ideal recipe is ultimately problem dependent. Carefully designing a PINN generally consists of identifying an optimal architecture to represent the PDE solutions, promoting a uniform loss landscape that contains the underlying physical system, and using an optimization algorithm that can minimize the loss function containing the PDE, boundary, and initial condition. 

Beginning with the architecture of the PINN, it will be taken to be a fully connected neural network that takes $(R,Z,t)$ as inputs, and one of the dependent variables $(\psi, g, u_R, u_Z)$ as an output. Thus, four networks will be considered, where we have found better performance in comparison to a single fully-connected neural network. We choose a network size of 5 hidden layers with 50 neurons in each layer for each neural network, and the $\tanh$ activation function.
\begin{figure}
\includegraphics[width=0.5\textwidth]{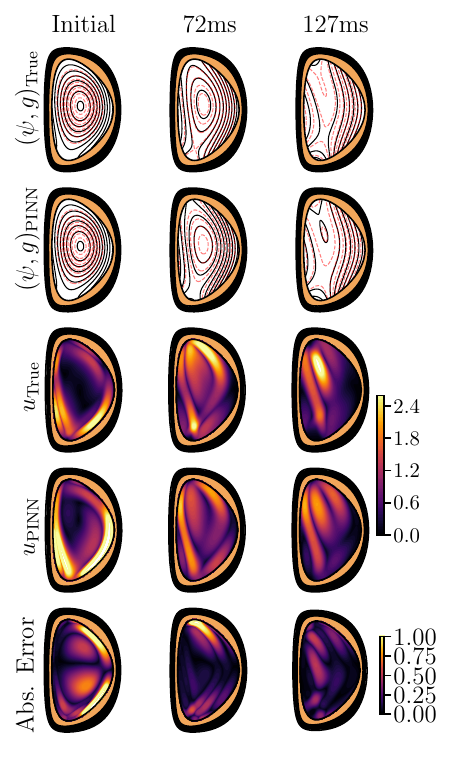}
\caption{Comparison PINN and true solutions.}
\label{fig:3}
\end{figure}
Moving to the loss function, which is taken to be
\begin{equation}
\mathcal{L} = \frac{w_\mathrm{IC}}{N_\mathrm{IC}}\sum_{i=0}^{N_\mathrm{IC}} \mathcal{I}^2 +  \frac{w_\mathrm{BC}}{N_\mathrm{BC}}\sum_{i=0}^{N_\mathrm{BC}} \mathcal{B}^2 +  \frac{w_\mathrm{PDE}}{N_\mathrm{PDE}}\sum_{i=0}^{N_\mathrm{PDE}} \mathcal{R}^2
, \label{eq:7}
\end{equation}
where $N_\mathrm{IC}, N_\mathrm{BC}, N_\mathrm{PDE}$ are the number of collocation points for the initial condition, boundary condition, and Equations~\ref{eq:3}-\ref{eq:6}, respectively, and $w_{(\cdot)}$ represents a weighting factor for each loss term. It is important to note that the continuous nature of neural networks will lead to difficulty in satisfying sharp transients, or discontinuities. We anticipate a sharp boundary layer as the flow velocity $\mathbf{u}$ will vanish at the vacuum chamber boundary, as well as sharp changes of the Lundquist number $S$ at the interface regions. As a result, the loss function will be peaked in these regions, leading to poor optimization and inaccurate solutions elsewhere in the domain. As a result, we will multiply the PDE residuals $\mathcal{R}$ by a function that rapidly vanishes as it approaches the boundaries and interfaces $\mathcal{R} \to \mathcal{R}[1.0 - \exp(\phi^2/\Delta\phi^2)]$, where $\phi$ is a distance function, and we choose $\Delta\phi = 0.025$. In addition, we note that under axisymmetry it can be shown that $\nabla\psi\cdot\mathbf{j} = (1/\mu_0)\nabla\varphi\cdot(\nabla g\times\nabla\psi) = 0$ in the vacuum chamber, which implies that $g = g(\psi)$. Hence, the additional constraint of $g$ being a function of $\psi$ can be incorporated into the training of the PINN, which will be done by including the mean squared error of $\nabla g\times \nabla\psi = 0$ as an additional loss term.

Turning to the optimization process that will minimize Equation~\ref{eq:7}, it is worth noting that conventional optimization routines employed in the broader machine learning literature are stochastic first order methods, such as the ADAM algorithm, which are attractive for its computational efficiency in speed and memory and its scalability to a large amount of GPUs. For PINNs, however, particularly in the limit of no synthetic or experimental data, first order methods are often not sufficient for finding the global minimum in non-convex landscapes. As a result, quasi-Newton methods are typically used to achieve better PINN performance, since they achieve a balance of providing curvature information of the loss landscape without directly computing the Hessian of the loss function. We thus use the self-scaled Broyden (SSBroyden) optimization algorithm, which has been shown to be extremely effective in achieving low levels of loss from its ability to maintain adequate curvature without becoming ill-conditioned~\cite{urban2025unveiling,kiyani2025optimizing,jnini2026curvature}. Regarding the collocation points, we shall allocate them in groups corresponding to the different regions in the geometry. We allocate half a million collocation points for both training and evaluation (test) across the entire time period in the vacuum chamber, followed by one hundred thousand collocation points each in the blanket and vessel regions. For the boundary condition collocation points, fifty thousand collocation points will be allocated. The collocation points are sampled with a quasi-Monte Carlo algorithm to prevent local groupings across the domain. For the weighting of each loss term, we weight the interface and boundary conditions terms less by 0.1,  we set the weights of Equations~\ref{eq:5} and ~\ref{eq:6} to 50, the weight for $\nabla g \times \nabla\psi$ to 10, and the rest are set to 1. During the training procedure, the collocation points will be adaptively resampled, where a larger portion of collocation points will be allocated where the magnitude of the loss is large~\cite{wu2023comprehensive}. Rather than use the total loss function in Equation~\ref{eq:7}, we only use the loss pertaining to Equations~\ref{eq:5} and \ref{eq:6}, as satisfying these equations is what gives the value of $\mathbf{u}$. Due to $\mathbf{u}$ only being implicitly defined, it is anticipated that the PINN will have the most difficulty in learning the correct solution for $\mathbf{u}$. Finally, The PINN will be trained on an Nvidia Blackwell 200 GPU (180 GB) and implemented with the PyTorch library in Python.

\begin{figure}
\includegraphics[width=0.5\textwidth]{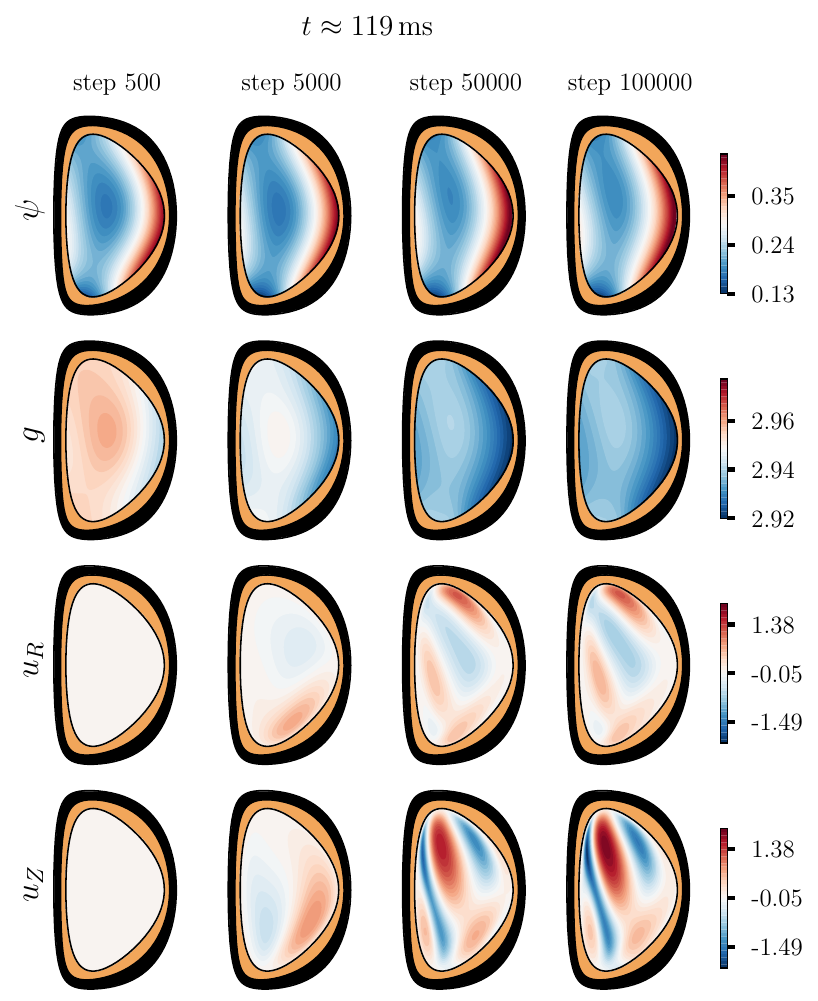}
\caption{Evolution of the predicted PINN solution during training.}
\label{fig:4}
\end{figure}

We now present the resulting calculation from the PINN. The training history of the PINN is shown in Figure~\ref{fig:2}, which occurred over the period of one day. Here, it can be seen that the loss decreases by roughly six to nine orders of magnitude, indicating that the PINN has learned a solution that generally satisfies the equations, and the occasional spikes in the training loss (solid curve) are from the adaptive resampling algorithm. Turning to the predicted solutions from the PINN, we choose an ion density of $10^{21}$ m$^{-3}$, which corresponds to a disrupted plasma that has undergone massive material injection, to compute the physical time. Verification of the PINN's solution is obtained by using the finite element method that uses curl-confirming and continuous Galerkin finite element spaces, solving Equations~\ref{eq:1} and ~\ref{eq:2} with a direct matrix solver, an adaptive implicit Runge-Kutta scheme for time integration,  and numerically implemented with the Firedrake library in Python~\cite{wimmer2024structure}. The resulting comparisons between the true and PINN solutions are shown in Figure~\ref{fig:3}, where the first two rows correspond to the scalar fields ($\psi$ in solid black, $g$ in dashed red), and the bottom two rows correspond to the magnitude of $\mathbf{u}$. We find that the PINN is able to learn a solution that satisfies $g = g(\psi)$ and generally agree with the true solution, where the error is largely at later times when the VDE is underway (127 ms). The error is mainly attributed to disagreement of the flow, which is shown in the on the bottom three rows. We find that the PINN was able to recover the general structure, however, the flow is under-predicted by the PINN during the VDE, and over predicted at the initial time. It is worth noting that while the loss history in Figure~\ref{fig:2} appears to be largely saturated, the monotonically decreasing curve implies that letting the PINN train for a longer period of time would yield a more accurate solution, which is shown in Figure~\ref{fig:4} by visualizing the PINN's solutions at a certain time throughout the training procedure. Indeed, we see that between steps 500 and 50,000 the PINN begins to learn the correct structure of $\mathbf{u}$, and the remaining training until step 100,000 is where the PINN learns the correct magnitude.

In summary, it is shown that a PINN can learn a set of time dependent MHD equations in tokamak geometry, which in the application of tokamak disruptions, can describe a vertical displacement event. With a modestly complex PINN design, all major qualitative trends were captured, where quantitative agreement improves as the PINN is trained for longer periods of time. Given the large amount of literature on variations of PINNs to tackle various problems~\cite{toscano2025pinns}, there is room to improve the results shown here, which we leave as future work. Noting that PINNs do not require discretizing the underlying domain, arbitrary geometries can be considered. Thus, implementing realistic tokamak geometry, such as the diverter region, is straightforward and subject to future work. In addition, learning the solutions over a broad range of plasma parameters (the Lundquist number for each region) can be done by including them as additional inputs to the PINN. While the training would be substantially longer, this remains an offline computation, where the online inference remains rapid. Given that no synthetic data was used during training, the encouraging results motivates future work to deploy operator learning methods, such as DeepONets~\cite{lu2021learning} and Fourier Neural Operators~\cite{li2020fourier} (see Ref.~\cite{carey2025neural} for an example in MHD) to learn the general operator for the MHD system that describes the VDE, where the physics-informed losses used in the PINN can also be included. As a result,  by training over both data from simulation and the governing equations, and a broad range of geometries and operating scenarios, the learned operator would serve as a high-fidelity surrogate that rapidly provides the plasma conditions during a VDE, and generalize to unseen scenarios. We also anticipate to interface the VDE surrogate in this work with other relevant physics models that are required to treat a disruption, such as the non-thermal plasma current carried by relativistic electrons, which has also received PINN treatment
~\cite{mcdevitt2023physics,mcdevitt2025physics,arnaud2025runaway}. As a result, combining both PINNs would enable an efficient means of assessing the damage on plasma facing components during tokamak disruption.

\section*{Acknowledgements}
This work was supported by the Department of Energy, Office of Fusion Energy Sciences at the University of Florida under award DE-SC0024649, at Los Alamos National Laboratory (LANL) under Contract No. 89233218CNA000001, and the Department of Energy, Office of Science Graduate Student Research (SCGSR) program administered by the Oak Ridge Institute for Science and Education under contract number DE‐SC0014664. This research used resources of the National Energy Research Scientific Computing Center (NERSC) using Award FES-ERCAP003229, and from University of Florida Research Computing (HiPerGator).

\bibliographystyle{apsrev}
\bibliography{./ref}

\end{document}